\documentclass[sigconf]{acmart}
\AtBeginDocument{%
  }

\usepackage{multirow}
\usepackage{float}
\usepackage{algorithmic}
\usepackage{textcomp}
\usepackage{xcolor}
\usepackage{listings}
\usepackage{bm}
\usepackage{enumitem}
\usepackage{makecell}

\begin{document}

\title{Secret Leak Detection in Software Issue Reports using LLMs: \\ A Comprehensive Evaluation}
% \title{Secret Breach Prevention in Software Issue Reports}

\author{Sadif Ahmed\textsuperscript{*}}
\email{ahmedsadif67@gmail.com}
\affiliation{%
  \institution{Bangladesh University of Engineering and Technology}
  %\department{Department of Computer Science and Engineering}
  \city{Dhaka}
  \country{Bangladesh}
}

\author{Md Nafiu Rahman\textsuperscript{*}}
\email{nafiu.rahman@gmail.com}
\affiliation{%
  \institution{Bangladesh University of Engineering and Technology}
  %\department{Department of Computer Science and Engineering}
  \city{Dhaka}
  \country{Bangladesh}
}

\author{Zahin Wahab\textsuperscript{*}}
\email{zahinwahab@gmail.com}
\affiliation{%
  \institution{The University of British Columbia}
  %\department{Department of Computer Science}
  \city{Vancouver}
  \state{BC}
  \country{Canada}
}

\author{Gias Uddin}
\email{guddin@yorku.ca}
\affiliation{%
  \institution{York University}
  %\department{Department of Electrical Engineering and Computer Science}
  \city{Toronto}
  \state{ON}
  \country{Canada}
}

\author{Rifat Shahriyar}
\email{rifat@cse.buet.ac.bd}
\affiliation{%
  \institution{Bangladesh University of Engineering and Technology}
  %\department{Department of Computer Science and Engineering}
  \city{Dhaka}
  \country{Bangladesh}
}

\thanks{\textsuperscript{*}Equal contribution. Author order does not matter}

% \author{Anonymous Author}

% \renewcommand{\shortauthors}{Wahab et al.}

\begin{abstract}
In the digital era, accidental exposure of sensitive information such as API keys, tokens, and credentials is a growing security threat. While most prior work focuses on detecting secrets in source code, leakage in software issue reports remains largely unexplored. This study fills that gap through a large-scale analysis and a practical detection pipeline for exposed secrets in GitHub issues. Our pipeline combines regular expression–based extraction with large language model (LLM)–based contextual classification to detect real secrets and reduce false positives. We build a benchmark of 54,148 instances from public GitHub issues, including 5,881 manually verified true secrets. Using this dataset, we evaluate entropy-based baselines and keyword heuristics used by prior secret detection tools, classical machine learning, deep learning, and LLM-based methods. Regex and entropy based approaches achieve high recall but poor precision, while smaller models such as RoBERTa and CodeBERT greatly improve performance (F1 = 92.70\%). Proprietary models like GPT-4o perform moderately in few-shot settings (F1 = 80.13\%), and fine-tuned open-source larger LLMs such as Qwen and LLaMA reach up to 94.49\% F1. Finally, we also validate our approach on 178 real-world GitHub repositories, achieving an F1-score of 81.6\% which demonstrates our approach’s strong ability to generalize to in-the-wild scenarios. 

\end{abstract}




\maketitle

\section{Introduction}
\label{sec:introduction}
In software systems, secrets act like digital keys that secure access to important resources and protect sensitive data. These include API keys, OAuth tokens, RSA encryption keys, TLS or SSL certificates, and user credentials such as usernames and passwords. As modern applications rely on many connected services, the chance of accidentally exposing these secrets has grown \cite{secretsaboutsecretsincode}. Platforms like GitHub and Bitbucket simplify development but also raise the risk of leaks. GitGuardian’s 2022 report \cite{secretsprawlreport} found over 10 million new hard-coded secrets, a 67\% increase from the previous year, across more than a billion scanned commits. Real-world cases show attackers exploiting exposed API keys in public repositories \cite{stealingAWScredentials}. A Sophos report also found that nearly half of all cyberattacks in early 2023 involved stolen credentials \cite{sophosreport}. Although cloud providers like Amazon and Google can detect and revoke compromised keys, version control systems still lack strong protection. Beyond source code, secrets can also appear in configuration files, test scripts, documentation, or logs \cite{secretsaboutsecretsincode}.

\begin{figure*}[h]
\centering
\includegraphics[width=0.7\textwidth]{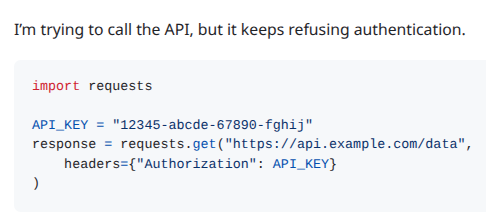}
\caption{Example of secret leak in GitHub issue report (Here we masked the actual key with a dummy key for security purposes)}
\label{fig:issue_example}
\end{figure*}

While most prior work has focused on detecting leaks in source code and configuration files, an overlooked but important risk comes from secrets unintentionally leaked in issue reports. Figure \ref{fig:issue_example} shows an example of this problem, where an API key is mistakenly shared in a public issue report.

Issue reports often include unstructured content such as logs, stack traces, shell commands, error messages, and example credentials that developers share while describing bugs or seeking help. Unlike source code, which follows a formal structure and consistent syntax, issue reports are written in natural language with a mixture of text, code snippets and screenshots. This lack of structure makes it much harder for automated tools to reliably identify whether a potential secret is real or just part of an example.

For instance, a developer might paste a failing API request such as:

\begin{quote} \texttt{curl https://api.stripe.com/v1/charges -u \\ sk\_test\_4eC39HqLyjWDarjtT1zdp7dc} \end{quote}

which unintentionally reveals a valid API key. Similarly, copied debugging logs can expose sensitive data like AWS credentials:

\begin{quote} \texttt{aws\_access\_key\_id=AKIAIOSFODNN7EXAMPLE} \end{quote}

On the other hand, issue reports may also contain obfuscated examples or placeholders intended for illustration, such as:

\begin{quote}
\texttt{apiKey=YOUR\_API\_KEY} \\
\texttt{token=sk\_test\_4eC39HqLyjWxxxxxxx}
\end{quote}

The coexistence of natural language, sample code, and actual credentials in issue reports makes secret detection far more complex than those in source code.

Traditional regex-based tools such as Gitleaks~\cite{gitleaks}, Gitrob~\cite{gitrob}, and TruffleHog~\cite{trufflehog2} rely on manually crafted regex patterns to identify secrets. However, even in source code, they often produce many false positives. To address these limitations, prior works used entropy-based filters~\cite{meli2019bad}, handcrafted feature-based classifiers~\cite{saha2020secrets}, and deep learning models~\cite{feng2022automated}. These achieve better precision in structured code environments. However, existing methods still fail to capture contextual information, making it hard to tell dummy keys from real secrets. Most have been developed and tested on source code rather than the more complex and unstructured issue reports. Large language models, with their strong grasp of context and ability to generalize across text types, present a promising solution. They can address these challenges without depending only on token patterns or entropy. Motivated by this, we investigate the use of LLM-based methods for detecting secrets in the unstructured and often noisy setting of GitHub issue reports.

In this work, we present the first large-scale study of secret leak detection in GitHub issue reports. We curated a benchmark dataset with 54,148 issues collected through the GitHub API. Our findings show that traditional methods like regex, entropy checks, and basic machine learning frameworks often flag too many false positives and fail to detect real secrets. In contrast, transformer models like RoBERTa and LLaMA use context to distinguish real secrets from random strings or placeholders. Our fine-tuned models significantly outperform older methods by improving precision through reducing false positives.

We explore this research through three primary research questions: 

\begin{itemize}

    \item \textbf{RQ1:} How do software developers perceive the likelihood, severity, and contributing factors of secret leaks in GitHub issue reports?
    \item \textbf{RQ2:} How effective are proprietary third-party large language models (LLMs) in detecting secrets within GitHub issue reports?
    \item \textbf{RQ3:} How effective are smaller, open-source Bert-like language models and LLMs in detecting secrets within GitHub issue reports?
\end{itemize}

In summary, our contributions are as follows:
\begin{enumerate}
    \item We have curated the first large-scale benchmark dataset for detecting secret leaks in software issue reports, containing over 54,000 labeled instances from diverse public GitHub repositories.
    \item We propose a robust secret detection pipeline that integrates regular expression-based extraction with contextual classification using LLMs 
    \item We conduct a comprehensive evaluation based on our proposed pipeline of both proprietary and open-source models, including smaller BERT-like models and larger LLMs like LLAMA and GPT, demonstrating that fine-tuned large language models (LLMs) achieve the best performance.  
    \item We evaluate our approach using 178 real-world GitHub repositories, demonstrating that the proposed models exhibit strong generalization capabilities beyond the curated dataset.  
  
\end{enumerate}

The rest of this paper is organized as follows. Section \ref{sec:motivation} presents the motivation and survey findings. Section \ref{sec:methodology} describes the dataset construction, preprocessing steps, and experimental setup. Section \ref{sec:evaluation} reports results for proprietary and open-source LLMs, and Section \ref{sec:ablation} presents ablation studies on learning rates and context sizes. Section \ref{sec:discussion} analyzes model behavior and deployment feasibility. Section \ref{sec:related} reviews related work, Section \ref{sec:threats} discusses validity threats, and Section \ref{sec:conclusion} concludes with future directions.

\textbf{Replication Package.} Our code and data are shared at 
\url{https://doi.org/10.5281/zenodo.17430335}

\section{Motivational Survey}
\label{sec:motivation}

To better understand why secrets appear in GitHub issue reports, we conducted a brief survey of software practitioners. While not statistically conclusive, their responses offered valuable practical insights that informed the direction of our subsequent technical analysis. In addressing \textbf{RQ1}, we focused on four central points:
(i) How likely do developers think secret leaks are in issue reports?
(ii) How serious do they see these leaks as being?
(iii) What kinds of situations tend to lead to accidental exposure?
(iv) What are the practices to detect secrets in issue reports?

\subsection{Survey Setup}

\paragraph{1) Survey Questions}
The survey included both quantitative and qualitative questions exploring the likelihood, severity, and causes of secret leaks in GitHub issue reports. The questionnaire was refined through discussions with software professionals and a pilot review to ensure clarity and relevance.

\paragraph{2) Survey Participants}
A total of \textbf{50 participants} took part in the survey. They were recruited through convenience sampling via the authors’ professional networks and online developer communities. The participants represented diverse professional roles and experience levels, as shown in Table~\ref{tab:participant_demographics}. Although convenience sampling introduces bias, we mitigated it by ensuring a balanced mix of developers, testers, team leads, and managers with varied years of experience.

\begin{table}[h]
\centering
\caption{Distribution of Survey Participants by Role and Experience.}
\label{tab:participant_demographics}
\resizebox{\columnwidth}{!}{%
\begin{tabular}{|l|c|c|c|c|c|}
\hline
\textbf{Role} & \textbf{0--2 yrs} & \textbf{3--5 yrs} & \textbf{6--10 yrs} & \textbf{10+ yrs} & \textbf{Total} \\ \hline
Developer & 12 & 8 & 3 & 2 & 25 \\ \hline
Tester & 6 & 3 & 2 & 1 & 12 \\ \hline
Team Lead & 2 & 2 & 1 & 1 & 6 \\ \hline
Other (Manager/Analyst) & 2 & 2 & 1 & 2 & 7 \\ \hline
\textbf{Total} & \textbf{22} & \textbf{15} & \textbf{7} & \textbf{6} & \textbf{50} \\ \hline
\end{tabular}%
}
\end{table}

\subsection{Likelihood of Secret Leaks}

Participants rated how likely they believe secret leaks are in GitHub issue reports on a 5-point scale (1 = least likely, 5 = most likely). Table~\ref{tab:rq1chart1} summarizes the distribution. More than 75\% (38 out of 50) rated the likelihood at 3 or higher, with six participants selecting the highest likelihood (5). These results suggest practitioners view secret leakage in issue reports as a recurring and credible risk, not a rare event.

\begin{table}[ht]
\centering
\small
\caption{Participants’ Ratings on the Likelihood of Secret Leakage in GitHub Issue Reports.}
\label{tab:rq1chart1}
\begin{tabular}{|c|c|c|}
\hline
\textbf{Rating} & \textbf{Number of Participants} & \textbf{Percentage (\%)} \\
\hline
1 (Least likely) & 0 & 0 \\
\hline
2 & 6 & 12 \\
\hline
3 & 23 & 46 \\
\hline
4 & 15 & 30 \\
\hline
5 (Most likely) & 6 & 12 \\
\hline
\textbf{Total} & 50 & 100 \\
\hline
\end{tabular}
\end{table}

\subsection{Perceived Severity of Secret Leaks}
Participants were further asked to assess the perceived severity of secret leaks within issue reports. As illustrated in Table~\ref{tab:rq1severity}, the majority (68\%) regarded such leaks as significant security threats, potentially leading to credential exposure or unauthorized system access. A smaller group (28\%) considered them moderately severe, while only a handful (4\%) viewed their impact as minimal.

\begin{table}[ht]
\centering
\small
\caption{Distribution of Responses on the Perceived Severity of Secret Leaks.}
\label{tab:rq1severity}
\begin{tabular}{|l|c|}
\hline
\textbf{Severity Category} & \textbf{Number of Respondents} \\ \hline
Highly severe & 34 \\ \hline
Moderately severe & 14 \\ \hline
Low severe & 2 \\ \hline
\textbf{Total} & \textbf{50} \\ \hline
\end{tabular}
\end{table}
The findings reveal a strong consensus among participants that the exposure of credentials constitutes a serious threat, carrying both security and organizational implications.

\subsection{Factors Behind Secret Sharing in Issue Reports}

We used open coding \cite{miles1994qualitative} and card sorting \cite{fincher2005making} to identify key factors. We gathered 45 responses, which led to 45 different codes grouped into four categories (see Table~\ref{tab:rq1factors}). The main reasons for secret exposure were a lack of awareness and time pressure during development.

\begin{table}[ht]
\centering
\small
\caption{Categories and Respondent Statistics for Factors Influencing Secret Sharing.}
\label{tab:rq1factors}
\begin{tabular}{|l|c|c|}
\hline
\textbf{Category} & \textbf{\#Codes} & \textbf{\#Respondents} \\ \hline
Lack of awareness or knowledge & 22 & 20 \\ \hline
Urgency or time pressure & 12 & 11 \\ \hline
Accidental sharing or oversight & 6 & 6 \\ \hline
Collaboration and team sharing & 5 & 5 \\ \hline
\textbf{Total} & \textbf{45} & \textbf{45} \\ \hline
\end{tabular}
\end{table}

Nearly half of the participants (44\%) pointed to \textbf{limited awareness or understanding of secure practices} as the primary reason for leaks. For example, one respondent noted, "not knowing what could be potential secrets". Urgency and time pressure (24\%) were also mentioned, as another respondent remarked, "It could be a time crunch/emergency situation/deadline that required hasty solution for a bug/problem". This is followed by accidental oversights (12\%) and team-based collaboration practices (10\%). 

\subsection{Detection Practices in Issue Reports}

To further support our findings, we asked participants how they usually find secrets in issue reports. As shown in Table \ref{tab:rq1detection}, most rely on manual checks or learn about leaks from others. None use automated tools, since most scanners target source code, not issue reports. This reveals a clear gap in tools for detecting secrets in issue reports.

\begin{table}[ht]
\centering
\small
\caption{Practices Used by Developers to Detect Secrets in Issue Reports.}
\label{tab:rq1detection}
\begin{tabular}{|l|c|}
\hline
\textbf{Detection Approach} & \textbf{Number of Respondents} \\ \hline
Manual inspection & 36 \\ \hline
Doesn't notice until someone points out & 14 \\ \hline
Automated tools & 0 \\ \hline
\textbf{Total} & \textbf{50} \\ \hline
\end{tabular}
\end{table}

Overall, the survey shows that developers see secret leaks in issue reports as common and serious. The lack of automated tools and reliance on manual checks point to the need for better detection methods.

\section{Methodology}
\label{sec:methodology} 
This section outlines the dataset construction, preprocessing, and modeling strategies employed to build and evaluate the proposed secret detection pipeline.
\subsection{Dataset}
\label{subsec:dataset}

% To the best of our knowledge, no prior work has investigated the detection of secret leaks within software issue reports. To support research in this direction, we curate and release a new benchmark dataset specifically designed for this task.
To the best of our knowledge, no prior work has investigated the detection of secret leaks in software issue reports. To support research in this direction, we curate and release a new benchmark dataset specifically designed for this task.

\paragraph{Collection of Issue Reports}
% We start by collecting issue reports from  public GitHub repositories using the REST API. Retrieval is guided by a large pool of targeted keyword queries such as \texttt{key}, \texttt{token}, \texttt{api key}, and \texttt{secret}, which help surface reports likely containing exposed secrets. For each retrieved issue, we retain the title, body, and repository metadata for downstream processing. Only active, non-forked repositories are included to ensure diversity and reproducibility.
Our data collection begins with retrieving issue reports from public GitHub repositories via the REST API. The retrieval process is driven by a broad set of targeted keywords such as \texttt{key}, \texttt{token}, \texttt{api key}, and \texttt{secret}, which help identify reports that are likely to contain exposed secrets. For each issue gathered, we preserve the title, body, and repository metadata to support subsequent analysis. To ensure both diversity and reproducibility, we include only active repositories that are not forks. The dataset spans a variety of domains, including web development, data science, DevOps, security tools, and mobile applications, providing a comprehensive view of secret leaks across different areas of software development.

\paragraph{Candidate Secret Extraction and Preprocessing}
% To find out if an issue actually contains a secret, we use 761 regular expressions used for secret detection in source code \cite{basak2023secretbench2}. But since issue reports aren't code, they often include things like commit hashes, UUIDs, build IDs, or random tokens that look like secrets but aren't. To cut down on false positives, we first filter out known harmless patterns using 22 regular expressions. After this cleanup, we scan the issue bodies with the secret-detection regexes and pull out each match as a candidate string. If there are multiple matches in one report, we keep them all to reflect real-world cases where several potential secrets might appear. Each final sample is a pair: {issue report body, candidate string}. The task is framed as binary classification, given a candidate string in an issue report, decide if it’s a real secret.

To determine whether an issue report actually contains a secret, we employ 761 regular expressions originally developed for secret detection in source code \cite{basak2023secretbench2}. However, since issue reports are not source code, they frequently include elements such as commit hashes, UUIDs, build IDs, or other random tokens that resemble secrets but are harmless. To reduce false positives, we first filter out known benign patterns using 22 additional regular expressions. Following this cleanup, we scan the issue bodies with the secret-detection regexes, extracting each match as a candidate string. When multiple matches occur in a single report, we retain all of them to capture realistic scenarios in which several potential secrets may coexist. Each resulting sample consists of a pair: {issue report body, candidate string}. We frame the task as a binary classification problem: given a candidate string within an issue report, the goal is to determine whether it constitutes a genuine secret.

\paragraph{Label Definition}
Each candidate string is manually annotated as either \textbf{Secret} (true positive) or \textbf{Non-sensitive} (false positive). Secrets refer to exposed credentials or cryptographic material such as API keys, tokens, private keys, or database connection strings with embedded credentials. Non-sensitives are strings that match detection patterns but do not represent real secrets, including placeholders (e.g., \texttt{YOUR\_API\_KEY}), redacted or masked values (e.g., \texttt{sk\_****************}) or \texttt{Bearer ey.\_<REDACTED>}) and dummy credentials (e.g., \texttt{apiKey=xxxxxxxxxx}).

\paragraph{Manual Annotation and Dataset Statistics}
All candidate strings were manually reviewed by the authors using consistent guidelines and considering the contextual meanings of the matched patterns to ensure accurate judgment. The manual verification relied on visual inspection and contextual reasoning rather than live credential testing. Annotators assessed how each candidate appeared in the issue report (e.g., in authentication steps, API requests, or configuration settings) and whether it was an example, placeholder, or masked value. Live validation against APIs or services was not performed due to ethical, legal, and security concerns. The final dataset contains \textbf{54{,}148} labeled instances, including \textbf{5{,}881} confirmed secrets and \textbf{48{,}297} non-secrets. These samples originate from \textbf{50{,}680} issue reports across \textbf{27{,}121} public repositories, covering a broad range of software ecosystems and languages. We provide the detailed annotation guidelines and examples used during manual review as part of our replication package.

\paragraph{Inter-Rater Agreement}
To measure annotation consistency, an additional author who did not participate in the original annotation process, independently labeled a random subset of \textbf{1{,}000} candidate strings. Table~\ref{tab:ira} shows the confusion matrix between the two annotators. The observed agreement is $P_o = 0.994$, and Cohen’s $\kappa$ \cite{cohen1960coefficient} is \textbf{0.9545}, indicating strong reliability.

\begin{table}[h]
\centering
\small
\caption{Inter-rater confusion matrix on 1{,}000 candidate strings (rows: Rater~2, columns: Rater~1).}
\begin{tabular}{|c|c|c|}
\hline
 & Rater 1: Positive & Rater 1: Negative \\
\hline
Rater 2: Positive & 68 & 4 \\
\hline
Rater 2: Negative & 2 & 926 \\
\hline
\end{tabular}
\label{tab:ira}
\end{table}

\paragraph{Train, Validation, and Test Splits}
We perform a split of \textbf{75\%/10\%/15\%} for training, validation, and testing, preserving class proportions. This ensures that rare secret cases remain evenly represented. Table~\ref{tab:split} shows the resulting distribution.

\begin{table}[h]
\centering
\small
\caption{Dataset split distribution.}
\label{tab:split}
\begin{tabular}{|l|c|c|c|}
\hline
\textbf{Split} & \textbf{Total} & \textbf{Secrets} & \textbf{Non-secrets} \\
\hline
Train (75\%)       & 40{,}593 & 4{,}377 & 36{,}216 \\
\hline
Validation (10\%)   & 5{,}432  & 586   & 4{,}846 \\
\hline
Test (15\%)         & 8{,}123  & 888   & 7{,}235 \\
\hline
\end{tabular}
\end{table}

\paragraph{Repository and Issue Characteristics}
We conducted a deeper analysis to understand where and how secret leaks arise across different repositories and issue types. At first, Table~\ref{tab:issue_type} presents the distribution of issue categories. Most leaks originate from \textbf{bug reports} (32.4\%), followed by \textbf{security-related} issues (20.5\%). A substantial portion (39.3\%) falls under miscellaneous categories labeled as “other,” indicating that secret exposures frequently emerge in smaller categories like routine maintenance or debugging discussions rather than bug or security related issues. The “Other” category primarily consists of unlabeled issues or issues tagged with minor or non standard labels such as environment setup, CI failures, maintenance tasks, or general discussions, which we group together to avoid introducing subjective re annotation bias.

\begin{table}[h]
\centering
\small
\caption{Distribution of secret leaks across issue types.}
\label{tab:issue_type}
\begin{tabular}{|l|c|}
\hline
\textbf{Issue Type} & \textbf{Percentage (\%)} \\
\hline
Other & 39.3 \\
\hline
Bug & 32.4 \\
\hline
Security & 20.5 \\
\hline
Feature & 5.0 \\
\hline
Question & 2.0 \\
\hline
Documentation & 0.8 \\
\hline
\end{tabular}
\end{table}

Table~\ref{tab:repo_size} reports the distribution of secret leaks across repositories of different sizes, together with the total number of repositories in each category. Notably, tiny repositories, despite being fewer in number (4,504), account for the highest number of leaks (1,396). In contrast, medium and large repositories are substantially more numerous (7,181 and 6,143 repositories, respectively) yet each exhibits a lower number of leaks (1,118 and 1,125). This numerical imbalance suggests that, on a per-repository basis, smaller projects are more prone to secret exposure. The trend likely reflects stronger governance, code review practices, and automated security scanning in larger projects, which help mitigate leaks despite their greater size and complexity.

\begin{table}[h]
\centering
\small
\caption{Distribution of repositories and secret leaks by repository size.}
\label{tab:repo_size}
\begin{tabular}{|l|c|c|}
\hline
\textbf{Repository Size} & \textbf{Total Repositories} & \textbf{Total Leaks} \\
\hline
Tiny   & 4,504 & 1,396 \\
\hline
Small  & 6,287 & 782 \\
\hline
Medium & 7,181 & 1,118 \\
\hline
Large  & 6,143 & 1,125 \\
\hline
Huge   & 2,617 & 835 \\
\hline
\end{tabular}
\end{table}

% \begin{table}[h]
% \centering
% \small
% \caption{Total secret leaks grouped by repository size.}
% \label{tab:repo_size}
% \begin{tabular}{|l|c|}
% \hline
% \textbf{Repository Size} & \textbf{Total Leaks} \\
% \hline
% Tiny & 1,396 \\
% \hline
% Small & 782 \\
% \hline
% Medium & 1,118 \\
% \hline
% Large & 1,125 \\
% \hline
% Huge & 835 \\
% \hline
% \end{tabular}
% \end{table}

\begin{figure}[h]
\centering
\includegraphics[width=\linewidth]{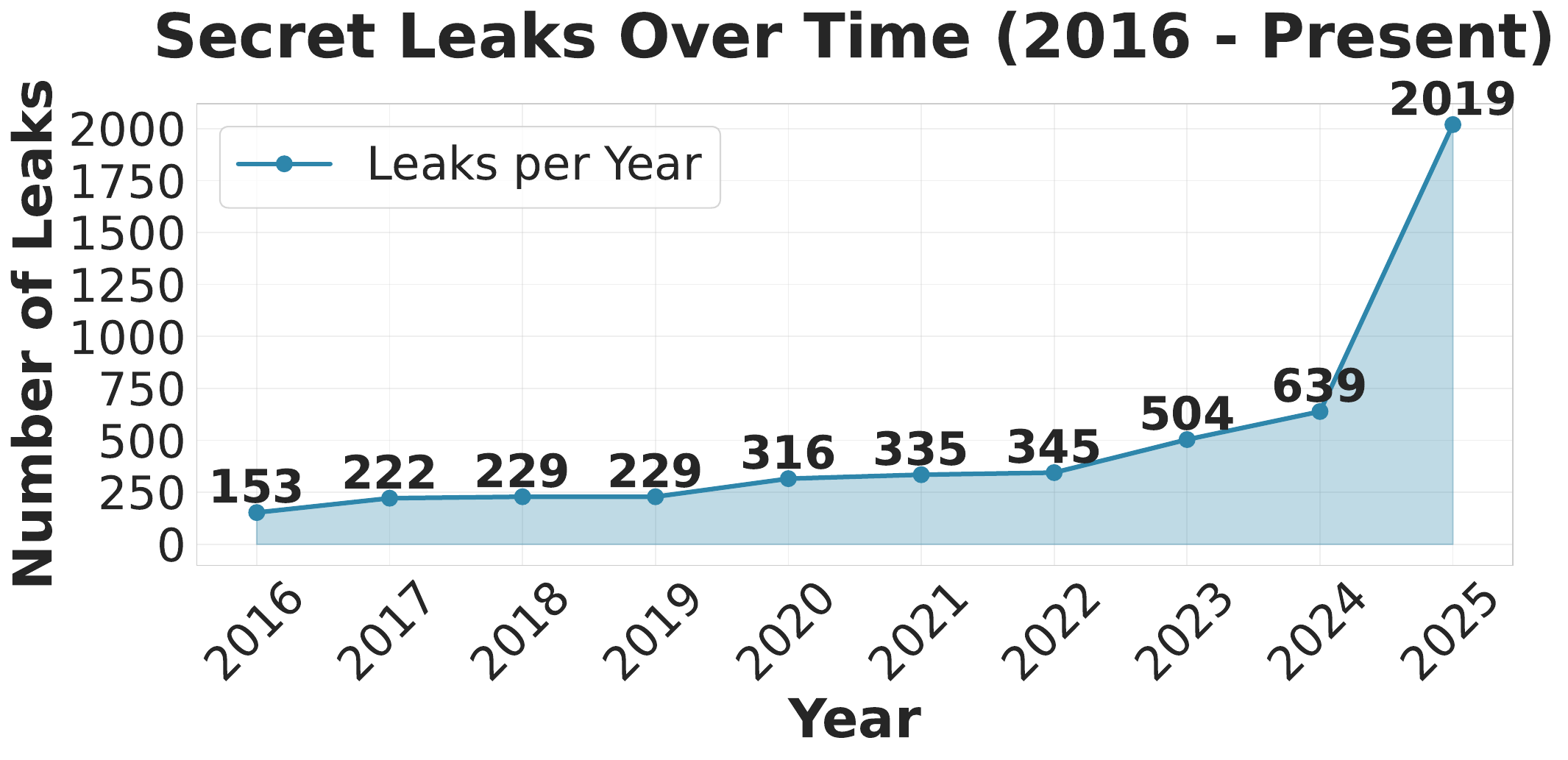}
\caption{Trend of secret leaks over time (2016–2025).}
\label{fig:time_trend}
\end{figure}

% Finally, Figure~\ref{fig:time_trend} illustrates how secret leaks have changed over time. Between 2016 and 2024, the number of leaks grew steadily but modestly, before rising sharply in 2025. The increase indicates growing use of APIs and cloud integrations in open-source projects and highlights the importance of automated leak detection in issue reports.
Finally, Figure~\ref{fig:time_trend} tells the story of the looming threats posed by secret leaks over the years. Between 2016 and 2024, leaks crept up steadily, almost unnoticed, before spiking sharply in 2025. This surge mirrors the growing reliance on APIs and cloud integrations in open-source projects and serves as a wake-up call for the urgent need for automated tools to detect secrets lurking in issue reports.

% Overall, these observations suggest that secret exposures in issue reports are not isolated or rare incidents. They occur across project sizes, appear in ordinary bug reports, and have become more frequent in recent years, underscoring the need for systematic detection and mitigation tools.
Overall, these findings indicate that secret exposures in issue reports are neither isolated nor uncommon. They span projects of all sizes, emerge even in routine bug reports, and have grown increasingly frequent in recent years, highlighting the pressing need for systematic tools to detect and mitigate such risks.
\subsection{Secret Detection Using Proprietary LLMs (RQ2)}
\label{subsec:proprietary_llm}
\textbf{Motivation:}
Pre-trained large language models (LLMs) have shown remarkable performance in text classification \cite{kostina2025largelanguagemodelstext} and contextual understanding \cite{zhu2024largelanguagemodelsunderstand} \cite{Wang2024AdaptableReliableTextClassification}, making them well-suited for secret detection. Their ability to capture contexts \cite{nguyen2024empiricalstudycapabilitylarge}, even without extensive task-specific fine-tuning, makes them promising for security-related applications. Motivated by this, we investigate the inherent capability of LLMs to identify secrets in issue reports. Specifically, we explore how well these models can distinguish true secrets from non-sensitive strings using contextual understanding without extensive fine-tuning. 

\textbf{Approach:}
In this study, we evaluated GPT-4o \cite{achiam2023gpt} and Gemini-2.0-Flash \cite{google2024gemini20flash} on detecting secrets and experimented with different prompting strategies such as zero-shot, and few-shot learning.

\textbf{Context Extraction and Data Preprocessing: }  Secret detection requires not only the candidate string but also its surrounding textual context within the issue report. Feng et al. showed that the relevant context lies within six lines of the secret \cite{feng2022automated}. To provide sufficient context while avoiding unnecessary overhead, we extract a \textbf{200-character window} around each candidate string. This local context helps the model capture usage patterns and distinguish between sensitive and benign content without processing the full issue report. As shown in our ablation study (Section \ref{sec:ablation}), a 200-character window provides the best balance between having enough context and keeping computation efficient, giving the model sufficient information for classification.

Each data point consists of the following: 

\begin{itemize}
    \item \textbf{Candidate String}: The string identified from the issue text using regex, which can or cannot be an actual secret.
    \item \textbf{Issue Context}: The 200-character excerpt surrounding the candidate string from the issue report.
    \item \textbf{Label}: A binary classification label indicating whether the candidate is a \textit{Secret} or \textit{Non-sensitive}.
\end{itemize}

\begin{figure}
    \centering
    \includegraphics[width=0.9\linewidth]{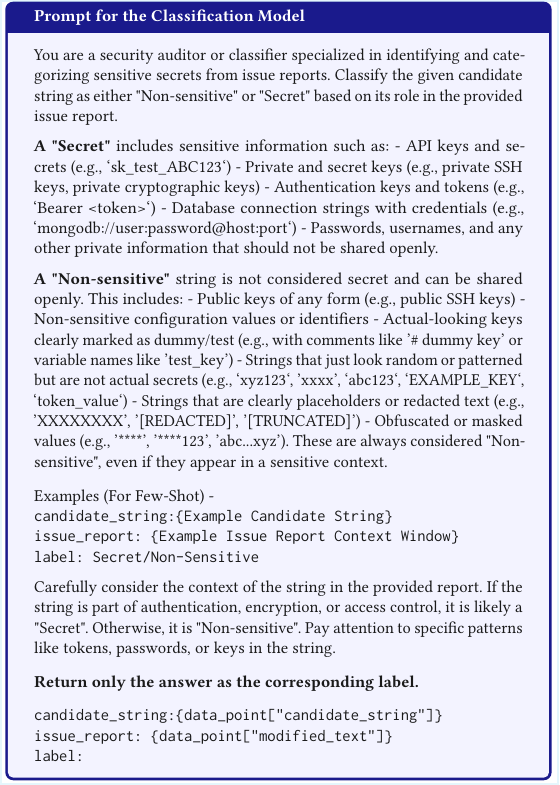}
    \caption{Prompt for the classification model.}
    \label{fig:prompt}
\end{figure}

\textbf{Prompt Engineering: } The extracted data is then transformed into structured natural language prompts to guide the model’s classification. The model is queried to determine whether a given candidate string within the issue context represents a secret. The prompt that we used is shown in Figure~\ref{fig:prompt}.

\paragraph{Zero-Shot Prompting}  
In the zero-shot setting, the model receives only the candidate string and its surrounding text, without any labeled examples. It relies solely on its pre-trained knowledge to determine whether the string is a secret. This setting serves as a baseline to assess the model’s inherent classification ability.

\paragraph{Few-Shot Prompting}  
Few-shot prompting \cite{brown2020language} extends the one-shot setting by including multiple labeled examples. These cover varied secret and non-secret cases, offering richer context. This setup improves model performance, especially on ambiguous inputs, by allowing it to learn from analogical reasoning.

\textbf{Parameter Settings: } To control the behavior of LLM outputs during inference, we adjust key generation parameters. The \texttt{temperature} parameter controls the randomness of the output distribution. Lower values make the model more deterministic, while higher values introduce more variability. We use a low \texttt{temperature} value (set to 0) to reduce randomness and ensure deterministic outputs.

\textbf{Evaluation Data: }  
We used the same dataset distribution as described in Section~\ref{subsec:dataset} for all experiments, ensuring consistency in evaluation across model families.

\textbf{Evaluation Metrics:}
We evaluate the models using the following standard classification metrics, each giving a different view of the model's performance: \\
\textbf{i) Precision:} It tells us what part of the model’s “secret” predictions are actually correct. A high precision means the model makes fewer mistakes and doesn’t trigger too many false security alerts. \\
\textbf{ii) Recall:} Recall measures how many actual secrets were correctly identified by the model. It captures the proportion of true positives among all actual positives. A high recall score ensures that most secrets are identified, reducing the risk of sensitive data exposure. \\
\textbf{iii) F1-Score:} The F1-score is the harmonic mean of precision and recall. It provides a balanced performance measure, especially useful when both false positives and false negatives need to be minimized.

\textbf{F1 Score for Balancing Precision and Recall:} In secret detection, it's important to catch as many real secrets as possible. While accuracy is useful during training, it doesn't show the balance between false positives and false negatives. To better measure model performance, we use the F1-score. It balances precision and recall, helping reduce both kinds of errors. This makes our evaluation more practical for security tasks, where too many alerts can overwhelm teams, and missed secrets can cause serious breaches.

\label{subsec:third_party_llm}
\subsection{Secret Detection Using Smaller, Open Source Language Models and LLMs (RQ3)}
\textbf{Motivation:}
Many organizations are cautious about using third-party large language models for software development due to compliance, data privacy, and high API costs. To address these concerns, we explore smaller, open-source LLMs, such as decoder-only architectures like LLaMA and Qwen, and encoder-based smaller language models like BERT. When fine-tuned on high-quality, domain-specific data, these models can approach the performance of larger LLMs while requiring far fewer resources, \cite{Alizadeh2023OpenSourceLLMsTextAnnotation} making them ideal for environments with limited hardware. Local deployment removes reliance on external APIs, eliminates variable usage costs, and ensures full control over sensitive data. This enhances security, supports compliance, and allows tighter integration with existing development workflows.

We further break down this research question into two subquestions:
\begin{itemize}[leftmargin=*, align=left, label={}, itemsep=1ex]
    \item \textbf{RQ3.1.} \hspace{1em}How effective are small Bert-like Language models in detecting secrets in GitHub issue reports?
    \item \textbf{RQ3.2.} \hspace{1em}How effective are open source LLMs in detecting secrets in GitHub issue reports? 
\end{itemize}

\subsubsection{Secret Detection with Small Bert-like Language Models}
\label{subsubsec:light_transformers}

 We fine-tuned the pre-trained lightweight transformer models such as RoBERTa-base \cite{Liu2019RoBERTa}, BERT-base-cased and BERT-base-uncased \cite{Devlin2019BERT}, ELECTRA-base \cite{Clark2020ELECTRA}, DistilBERT (cased and uncased) \cite{Sanh2019DistilBERT}, CodeBERT-base \cite{Feng2020CodeBERT}, ALBERT-base-v2 \cite{Lan2020ALBERT}, XLNet-base-cased \cite{Yang2019XLNet}, BigBird-RoBERTa-base \cite{zaheer2021bigbirdtransformerslonger}, Funnel-Transformer \cite{Dai2020FunnelTransformer}, and LUKE-base \cite{Yamada2020LUKE} to perform binary classification for secret detection, as these models give strong classification performance \cite{Fadel2024PerformanceEvaluationBERTTextClassification}. Unlike large language models that rely on structured prompt-based interactions, these models directly consume fixed-format input sequences. 

\textbf{Context Extraction and Data Preprocessing: }  
Each training instance comprises the candidate string together with a 200-character context window extracted from the issue report, following the same context extraction process described in Section~\ref{subsec:proprietary_llm}.

The data points are constructed in an identical way as Section \ref{subsec:proprietary_llm}

\textbf{Input Representation and Training: }  
Each sample is represented as a flat text sequence containing the extracted context window and candidate string. The sequence is tokenized and encoded using the respective transformer model. A fully connected classification head is added on top of the [CLS] token (or equivalent pooled representation) to predict the binary label (\textit{Secret} or \textit{Non-sensitive}).  
We use cross-entropy loss for optimization, the AdamW optimizer with a learning rate of $2\times 10^{-5}$, and early stopping based on validation loss to prevent overfitting. Fine-tuning was conducted for 10 epochs with a batch size of 8.

\subsubsection{Secret Detection with open source LLMs}
\label{subsubsec:small_llm_detection}

We fine-tuned several LLMs such as LLaMA-3.1-8B, Mistral-7B, DeepSeek-7B, Qwen-7B and Gemma-7B to detect secrets in issue reports. These models were selected for their strong performance on complex language tasks and efficient inference capabilities. We fine-tuned these models for secret detection using structured prompts.

\textbf{Context Extraction and Data Preprocessing: }  
Each training instance comprises the candidate string together with a \textbf{200-character context window} extracted from the issue report, following the same context extraction process described in Section~\ref{subsec:proprietary_llm}.

\textbf{Prompt Engineering: }  
We used the same prompt format outlined in Section \ref{subsec:proprietary_llm} for the fine-tuned models constructing data points in an identical way.

\textbf{Parameter Settings: }  
For fine-tuning, we used \textbf{QLoRA} \cite{dettmers2023qloraefficientfinetuningquantized} with 4-bit precision (NF4 quantization \cite{yoshida2023nf4isntinformationtheoretically} and FP16 computation \cite{micikevicius2018mixedprecisiontraining}), disabling double quantization to reduce memory usage. Fine-tuning focused on transformer attention layers for domain-specific adaptation. We employed parameter-efficient fine-tuning (PEFT) \cite{xu2023parameterefficientfinetuningmethodspretrained} using LoRA \cite{dettmers2023qloraefficientfinetuningquantized} with rank 64 and $\alpha=16$, without dropout or bias adjustments. The \texttt{Paged AdamW} optimizer \cite{bitsandbytes_AdamW} (learning rate $2\times10^{-4}$) was used with memory-efficient paging. Training ran for 5 epochs with batch size 1, 8 gradient accumulation steps, and a cosine decay scheduler with 3\% warm-up.

\textbf{Evaluation Data: }  
We used the same data distribution as described in Section~\ref{subsec:dataset} for all experiments.

\begin{figure*}[t]
    \centering
    \includegraphics[width=0.8\linewidth]{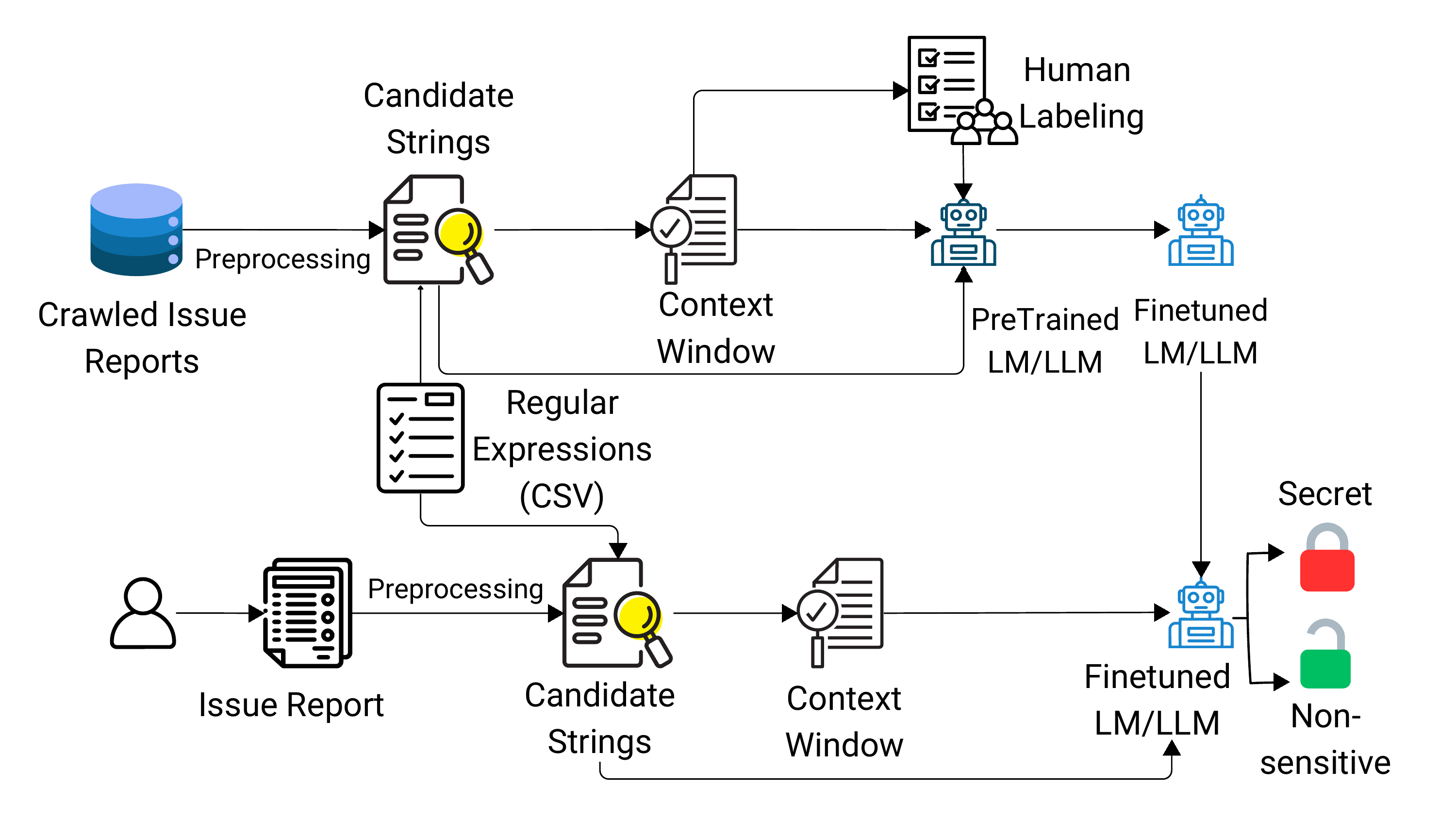} 
    \captionsetup{font=normalsize}
    \caption{Workflow for secret detection in issue reports. Candidate strings are extracted using 761 regular expressions and a 200-character context window is created around each. Human-labeled samples are used to fine-tune a language model during training. In inference, the same extraction and context generation steps are applied, and the fine-tuned model classifies each candidate as secret or non-sensitive.}

    \label{pipeline}
\end{figure*}

Figure~\ref{pipeline} illustrates the overall workflow of our proposed secret detection system.

\section{Evaluation}
\label{sec:evaluation}

This section presents an empirical evaluation of our proposed secret detection method, comparing proprietary and open-source models across multiple settings.

\subsection{RQ2: Effectiveness of Proprietary LLMs}
To answer RQ2, we evaluate GPT-4o and Gemini-2.0-flash under zero-shot and few-shot prompting to see how well these large models can detect secrets without fine-tuning. Table~\ref{llm-results} presents the results in terms of precision, recall, and F1-score.

\begin{table}[h]
\centering
\small
\caption{Evaluation Metrics for Zero-shot and Few-shot Learning on Pretrained LLMs}
\label{llm-results}
\begin{tabular}{|p{1.2cm}|c|p{1cm}|p{0.8cm}|p{0.8cm}|}
\hline
\textbf{Setting} & \textbf{Model} & \textbf{Precision} & \textbf{Recall} & \textbf{F1-Score}  \\
\hline
\multirow{2}{*}{Zero-Shot} 
& Gemini-2.0-flash & 44.96\% & 85.92\% & 59.03\% \\
& GPT-4o & 59.25\% & 90.67\% & 71.66\% \\
\hline
\multirow{2}{*}{Few-Shot} 
& Gemini-2.0-flash & 55.30\% & 86.94\% & 67.60\% \\
& GPT-4o & \textbf{70.47\%} & \textbf{92.86\%} & \textbf{80.13\%} \\
\hline
\end{tabular}
\end{table}

From Table~\ref{llm-results}, we can see that GPT-4o performs better overall than Gemini-2.0-flash in both zero-shot and few-shot settings. While Gemini tends to achieve higher recall, GPT-4o consistently shows better balance across all metrics.

\begin{itemize}
\item \textbf{Zero-Shot:} Gemini-2.0-flash shows strong sensitivity with 85.92\% recall but lower precision (44.96\%), indicating high false positives. GPT-4o performs better overall with higher precision (59.25\%), recall (85.92\%) and F1-score (71.66\%).

\item \textbf{Few-Shot:} When given a few examples, both models improve, but GPT-4o shows a clear lead. It reaches 70.47\% precision, 92.86\% recall, and 80.13\% F1-score, indicating stronger performance compared to Gemini-2.0-flash.
\end{itemize}

Overall, among the proprietary models, GPT-4o proves to be the more effective and reliable model for secret detection in issue reports, offering higher accuracy and better generalization in both prompting setups.

\subsection{RQ3: Effectiveness of Smaller, Open-Source LLMs}

\paragraph{RQ3.1: Effectiveness of Small Bert-like Language Models}

Table~\ref{tab:transformer-results} summarizes the results for lightweight transformer models such as BERT, RoBERTa, and CodeBERT. These models achieve strong performance, with F1-scores for the positive class (Class 1) between \textbf{90\% and 93\%}.

\begin{table}[htbp]
  \centering
  \caption{Per-model results: Small Bert-like LLMs}
  \label{tab:transformer-results}
  \small
  \begin{tabular}{|l|c|c|c|}
    \hline
    \textbf{Model} & \textbf{Precision} & \textbf{Recall} & \textbf{F1-Score} \\
    \hline
    RoBERTa-base & 93.03\% & 91.67\% & 92.34\% \\
    BERT-base-cased & 92.87\% & 89.53\% & 91.17\% \\
    BERT-base-uncased & 91.09\% & 92.12\% & 91.60\% \\
    ELECTRA-base & 91.24\% & 90.32\% & 90.78\% \\
    DistilBERT-cased & 91.76\% & 87.84\% & 89.76\% \\
    CodeBERT-base & \textbf{92.49\%} & 92.91\% & \textbf{92.70\%} \\
    DistilBERT-uncased & 89.68\% & 90.99\% & 90.33\% \\
    ALBERT-base-v2 & 82.67\% & 85.92\% & 84.26\% \\
    XLNet-base-cased & 88.92\% & \textbf{94.93\%} & 91.83\% \\
    BigBird-RoBERTa-base & 83.17\% & 87.39\% & 85.23\% \\
    Funnel-Transformer & 72.60\% & 89.53\% & 80.18\% \\
    LUKE-base & 92.44\% & 92.23\% & 92.33\% \\
    \hline
  \end{tabular}
\end{table}

The results show that smaller BERT-like models perform very well for this task. Most models reach F1-scores above 90\% for the positive class (Secret). Among them, \textbf{CodeBERT-base} gives the best overall performance with the highest F1-score (\textbf{92.70\%}), combining strong precision (92.49\%) and recall (92.91\%). \textbf{XLNet-base-cased} achieves the best recall (\textbf{94.93\%}), meaning it is better at avoiding missed secrets, although its precision (88.92\%) is a bit lower. On the other hand, smaller or less common models such as \textbf{ALBERT-base-v2} and \textbf{Funnel-Transformer} falls behind, with F1-scores around 84.3\% and 80.2\%, indicating a trade-off between model complexity and performance. Overall, these models show strong results while remaining efficient, which makes them suitable for resource constrained environment.

\paragraph{RQ3.2: Effectiveness of open source LLMs}
We now turn to fine-tuned LLaMA, Mistral, DeepSeek, Qwen, and Gemma. Table~\ref{tab:llm-results} reports their performance. These models consistently outperform lightweight transformers, reaching F1-scores to 94\%.

\begin{table}[htbp]
  \centering
  \caption{Per-model results: Fine-tuned Larger LLMs}
  \label{tab:llm-results}
  \small
  \begin{tabular}{|l|c|c|c|}
    \hline
    \textbf{Model} & \textbf{Precision} & \textbf{Recall} & \textbf{F1-Score} \\
    \hline
    LLaMA-3.1-8B  & \textbf{95.89\%} & 92.00\% & 93.91\% \\
    Mistral-7B    & 94.28\% & 90.88\% & 92.55\% \\
    DeepSeek-7B   & 92.55\% & 93.45\% & 93.00\% \\
    Qwen-7B       & 94.78\% & \textbf{94.14\%} & \textbf{94.46\%} \\
    Gemma-7B      & 93.24\% & 94.33\% & 93.78\% \\
    \hline
  \end{tabular}
\end{table}

\noindent
LLaMA-3.1-8B exhibits the highest precision among the models at \textbf{95.89\%}, while \textbf{Qwen-7B} delivers the best overall performance, achieving a strong precision of 94.78\%, the highest recall of \textbf{94.14\%}, and the top F1-score of \textbf{94.46\%}, making it the most balanced and effective model across all metrics. The remaining models like Mistral-7B, DeepSeek-7B, and Gemma-7B also show competitive performance, with F1-scores ranging between 92.55\% and 93.78\%.

\begin{figure}[h]
    \centering
    \includegraphics[width=0.5\textwidth]{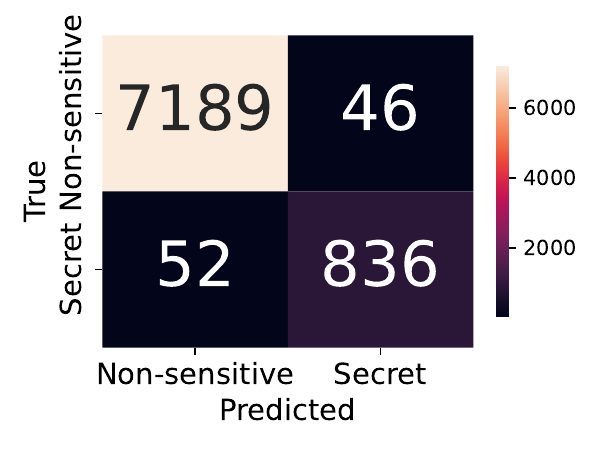}
    \caption{Qwen-7B confusion matrix for secret detection.}
    \label{fig:confusion-matrix}
\end{figure}

To illustrate the detection dynamics more clearly, Figure ~\ref{fig:confusion-matrix} shows the confusion matrix for Qwen-7B. The matrix shows that out of 7,235 actual non-secrets, the model correctly rejected 7,189 while incorrectly flagging only 46 as secrets. Similarly, out of 888 actual secrets, 836 were correctly identified, with just 52 missed. This balance between false positives and false negatives explains the model’s high precision and recall.

We compared our transformer and large language model approaches with traditional baselines grouped into three types: regex and heuristic methods, classical machine learning models, and deep learning architectures. For the baselines, we used a mix of regex and entropy-based heuristics following \citet{meli2019bad}. We also combined regex methods with the criteria mentioned in our prompt (Figure \ref{fig:prompt}). We then trained feature-based classifiers such as Decision Tree, Random Forest, K-Nearest Neighbors, Logistic Regression, Support Vector Machine, and Naive Bayes as described by \citet{saha2020secrets}. Some features from the original work relied on source code metadata, which were unavailable for issue reports, so we focused on textual and entropy-based features. For the deep learning baseline, we implemented a TextCNN model adapted from \citet{feng2022automated}. The original network was designed for three-class password classification, and we adjusted it for binary secret detection while keeping its architecture and parameters unchanged. As shown in Table~\ref{tab:summary-contrast}, transformer and LLM models clearly outperform all baselines. Regex methods achieve high recall but poor precision, often labeling everything as sensitive. Even after incorporating placeholder, masking, and dummy-value heuristics along with regex, the precision stays low due to lack of generalizability (37.66\%). Machine learning classifiers perform better but remain limited by handcrafted features and lack of context. TextCNN improves further with an F1-score of 84.19\%, yet Bert-like models exceed 92\%, and fine-tuned larger LLMs reach over 94\%. These results confirm that contextual information and large-scale pretraining are key for detecting secrets in noisy issue report text.

\begin{table}[htbp]
  \centering
  \small
  \caption{Summary comparison: Regex, ML, and DL baselines vs. Transformer models vs. Large Language Models.}
  \label{tab:summary-contrast}
  \resizebox{\columnwidth}{!}{%
    \begin{tabular}{|l|c|c|c|}
      \hline
      \textbf{Method} & \textbf{Precision} & \textbf{Recall} & \textbf{F1-Score} \\
      \hline
      Regex & 6.8\% & 100.0\% & 12.8\% \\
      Regex + Entropy & 11.2\% & 97.5\% & 20.1\% \\
      Regex + Entropy + Keyword + Heuristic & 13.0\% & 68.7\% & 21.8\% \\
      Regex + Criteria in The Prompt & 25.39\% & 72.86\% & 37.66\% \\
      Decision Tree & 69.67\% & 47.07\% & 56.18\% \\
      Random Forest & 74.41\% & 46.51\% & 57.24\% \\
      K-Nearest Neighbors & 66.91\% & 51.46\% & 58.18\% \\
      Logistic Regression & 76.74\% & 40.88\% & 53.34\% \\
      SVM & 77.75\% & 38.96\% & 51.91\% \\
      Naive Bayes & 25.24\% & 83.00\% & 38.71\% \\
      TextCNN & 93.66\% & 76.46\% & 84.19\% \\
      \textbf{Small Bert-like model} & 92.49\% & 92.91\% & 92.70\% \\
      \textbf{LLM} & 94.78\% & 94.14\% & 94.46\% \\
      \hline
    \end{tabular}
  }
\end{table}

\section{Ablation Study}
\label{sec:ablation}

To better understand how configuration choices influence performance, we conduct an ablation study focusing on two critical factors: the learning rate used during fine-tuning and the size of the context window provided to the model. Experiments were performed on the held-out test set using two representative models, \textbf{CodeBERT-base} (transformer family) and \textbf{Qwen-7B} (fine-tuned LLM), since they are the best-performing models from RQ3.1 and RQ3.2 respectively.

\subsection{Impact of Learning Rate}
\label{subsec:ablation-lr}

Learning rate (LR) directly affects convergence behavior during fine-tuning. Table~\ref{tab:lr-ablation} reports how varying the LR influences classification accuracy for both CodeBERT and Qwen. Extremely low values slow down convergence and underfit the decision boundary, while overly high values destabilize training and degrade precision. A moderate LR yields the best balance between recall and precision for both models.

\begin{table}[h]
\centering
\small
\caption{Effect of Learning Rate on Model Performance}
\label{tab:lr-ablation}
\resizebox{\columnwidth}{!}{%
\begin{tabular}{|l|c|c|c|c|}
\hline
\textbf{Model} & \textbf{Learning Rate} & \textbf{Precision} & \textbf{Recall} & \textbf{F1-Score} \\
\hline
\multirow{3}{*}{CodeBERT-base} 
& $1\times10^{-5}$ & 91.20\% & 92.50\% & 91.85\% \\
& $2\times10^{-5}$ & \textbf{92.49\%} & \textbf{92.91\%} & \textbf{92.70\%} \\
& $4\times10^{-5}$ & 90.52\% & 91.84\% & 91.10\% \\
\hline
\multirow{3}{*}{Qwen-7B} 
& $1\times10^{-4}$ & 93.60\% & 93.80\% & 93.70\% \\
& $2\times10^{-4}$ & \textbf{94.78\%} & \textbf{94.14\%} & \textbf{94.46\%} \\
& $4\times10^{-4}$ & 93.05\% & 93.28\% & 93.17\% \\
\hline
\end{tabular}
}
\end{table}

Both models achieve optimal results at moderate learning rates—$2\times10^{-5}$ for CodeBERT and $2\times10^{-4}$ for Qwen, indicating consistent convergence stability across scales. The Qwen model exhibits slightly higher precision and recall overall, reflecting its stronger contextual adaptation during optimization.

\subsection{Impact of Context Window Length}
\label{subsec:ablation-context}

The context window refers to the amount of text the model considers around the candidate string during classification. A smaller window might miss important context, while a larger window can increase training and inference time and use more memory. Table~\ref{tab:context-ablation} shows the performance across three different window sizes.

\begin{table}[h]
\centering
\small
\caption{Effect of Context Window Length on Model Performance}
\label{tab:context-ablation}
\resizebox{\columnwidth}{!}{%
\begin{tabular}{|l|c|c|c|c|}
\hline
\textbf{Model} & \textbf{Context (chars)} & \textbf{Precision} & \textbf{Recall} & \textbf{F1-Score} \\
\hline
\multirow{3}{*}{CodeBERT-base} 
& 100 & 91.80\% & 92.10\% & 91.95\% \\
& 200 & \textbf{92.49\%} & \textbf{92.91\%} & \textbf{92.70\%} \\
& 300 & 92.60\% & 92.98\% & 92.79\% \\
\hline
\multirow{3}{*}{Qwen-7B} 
& 100 & 94.00\% & 93.80\% & 93.90\% \\
& 200 & \textbf{94.78\%} & \textbf{94.14\%} & \textbf{94.46\%} \\
& 300 & 94.90\% & 94.20\% & 94.55\% \\
\hline
\end{tabular}%
}
\end{table}

Both models exhibit steady improvement as the window expands from 100 to 200 characters, reflecting enhanced contextual understanding of longer keys and embedded variables. However, gains beyond 400 characters are marginal, suggesting that moderate window sizes suffice for most practical cases.

\section{Discussion}
\label{sec:discussion}

This section summarizes insights drawn from the experimental results, highlighting model behavior, strengths, limitations and performance observed in real-world scenarios.

\subsection{Analysis of Detected Secrets}
To evaluate the fine-tuned model performance in secret detection, we evaluate its ability to correctly distinguish real secrets from false positives in issue reports. We also examine the cases where the model misclassifies secrets to better understand its limitations.

\paragraph{True Positives} The model successfully identified genuine sensitive information embedded within real-world issue report contexts. For example, in one case, a developer shared a Stripe test API key in an issue while reporting a failed payment request:
\begin{quote}
“The payment keeps failing with a 401 error. Here's the request I'm using: POST /v1/charges with header \emph{Authorization: Bearer sk\_test\_4eC39HqLyjWDarjtT1zdp7dc}.”
\end{quote}

This example demonstrates the model’s capability to accurately detect secrets embedded in noisy, unstructured contexts such as API traces and debugging notes.

\paragraph{False Positives} The model also showed strong performance in minimizing false positives, accurately recognizing non-sensitive content even when it closely resembled secrets. For instance, a user reporting a login issue wrote:
\begin{quote}
“Trying to sign in with user: \emph{user123} and \emph{pass: password} – login keeps failing on test server.”
\end{quote}
Although the use of hardcoded strings could trigger regex-based tools, the model correctly ignored this as a non-secret, likely due to its low entropy and placeholder-like nature.

Another issue included a mocked API key:
\begin{quote}
“To reproduce the bug, you can use dummy credentials like \emph{apiKey=xxxxxxxxxx} and \emph{token=your\_dummy\_key}.”
\end{quote}
The model accurately detected that such entries were synthetic placeholders and not real secrets, avoiding unnecessary false alarms.
These results highlight the model's strength in contextual judgment, balancing precision and recall.

\subsection{Discussion on False Negatives and Misclassified Cases}
\label{subsec:false-negatives}

While our fine-tuned models achieved high overall precision and recall, certain edge cases exposed inherent challenges in secret classification. In this subsection, we analyze representative false negatives and ambiguous samples that reveal nuanced limitations in contextual interpretation.
\paragraph{Ambiguous Placeholder Passwords.} 
One class of misclassification arises from passwords that appear syntactically sensitive but are semantically safe. For example:
\begin{verbatim}
AIRFLOW_HOME=/opt/airflow
AIRFLOW_USER=admin
AIRFLOW_PASSWORD=airflow
\end{verbatim}
In this case, annotators labeled the value \texttt{airflow} as sensitive out of caution, but the model identified it as non-sensitive. This illustrates how labeling subjectivity can sometimes conflict with contextual understanding. Here, “airflow” refers to Apache Airflow, a well-known open-source platform. Such names are commonly used as a placeholder or default credential in example configurations. Such values are often publicly known and not intended to be secret. Whether these strings are considered sensitive depends heavily on the user's intent and deployment context, making them inherently ambiguous.
\paragraph{Truncated or Partial Key Contexts.} 
Another observed source of false negatives occurred in long cryptographic keys, such as RSA private keys, where later portions of the key were truncated or masked (e.g., with ``\texttt{****}'', ``\texttt{...}'', or ``\texttt{[truncated]}'' markers). When the model's context window was too short, these truncated indicators sometimes fell outside the captured segment, leading the classifier to treat the sample as benign. Increasing the contextual window length restored detection accuracy, confirming that long-range dependencies play an important role in such cases.
\paragraph{Inherent limitations of LLMs.} A small number of missed detections reflect the inherent limitations of current language models. Some errors stem not from label ambiguity or missing context, but from inconsistencies in semantic understanding, showing that even advanced models can behave unpredictably with subtle linguistic cues or domain-specific semantics.

\subsection{Effectiveness of Model on Real World Repositories}

To evaluate how well the model generalizes in real-world conditions, we evaluated it on 178 public GitHub repositories covering different areas such as cloud infrastructure (terraform provider aws), API clients (google api nodejs client), web platforms (vercel), and data systems (redis). Out of 1,489 issue reports, 30 were actual secret cases. Despite this strong class imbalance, the model achieved a macro average F1 score of 81.60\%. For the Secret class, it reached a precision of 50.98\% and a recall of 86.67\%, showing that it can catch most real secrets while producing a moderate number of false positives. As shown in Table~\ref{tab:wildresult}, the macro average metrics confirm balanced performance across classes, and the confusion matrix in Table~\ref{tab:wildconf} highlights the model’s high recall. The full list of the 178 repositories is available in the replication package.

\begin{table}[h]
\caption{Performance metrics for large-scale evaluation on 178 real-world repositories}
\centering
\small
\setlength{\tabcolsep}{6pt}
\begin{tabular}{|c|c|c|c|c|}
\hline
\textbf{Class} & \textbf{Precision} & \textbf{Recall} & \textbf{F1-score} \\
\hline
\textbf{Secret} & 50.98\% & 86.67\% & 64.20\% \\
\textbf{Macro Avg.} & 75.35\% & 92.45\% & 81.60\%  \\
\hline
\end{tabular}
\label{tab:wildresult}
\end{table}

\begin{table}[h]
\caption{Confusion matrix for large-scale evaluation}
\centering
\small
\setlength{\tabcolsep}{12pt}
\begin{tabular}{|c|c|c|}
\hline
 & \makecell[c]{\textbf{Predicted} \\ \textbf{Non-sensitive}} & \makecell[c]{\textbf{Predicted} \\ \textbf{Secret}} \\
\hline
\textbf{Actual Non-sensitive} & 1434 & 25 \\
\textbf{Actual Secret} & 4 & 26 \\
\hline
\end{tabular}
\label{tab:wildconf}
\end{table}

\subsection{Training Setup}
All fine-tuning was done on a local workstation with an Intel Core i5-13400F, 128 GB RAM, and an RTX 4090 GPU. Using QLoRA, PEFT, and Unsloth, we kept GPU memory usage for 7B–8B models within 7--8 GB. This shows large-scale fine-tuning is feasible on high-end consumer hardware without cloud resources.

\subsection{Training and Inference Efficiency}
We evaluated the resource usage and performance of both small and large fine-tuned LLMs. Training remained feasible within 6–7 GB of VRAM for small BERT-like models and 8–10 GB for larger LLMs. Corresponding training times averaged 2–3 hours and 14–16 hours, respectively. During inference, memory usage stayed below 6 GB across all cases, with per-sample latency ranging from 0.05 to 0.13 seconds. These results demonstrate that even large fine-tuned LLMs can be efficiently integrated into developer environments without incurring significant computational overhead.

\begin{table}[h]
\centering
\small
\caption{Resource Efficiency Metrics}
\label{tab:resource-efficiency}

\begin{tabular}{|p{1.8cm}|p{0.8cm}|p{0.9cm}|p{1.2cm}|p{1.2cm}|}
\hline
\textbf{Model Family} & \textbf{Train VRAM (GB)} & \textbf{Train Time (hrs)} & \textbf{Inference VRAM (GB)} & \textbf{Inference Time (s)} \\
\hline
Small Bert-like models  & 6-7 & 2-3 & 3-4 & 0.05--0.07 \\
Larger LLMs  & 8-10 & 14-16 & 5-6 & 0.10--0.13 \\
\hline
\end{tabular}
\end{table}

\section{Related Work}
\label{sec:related}

\paragraph{Large-Scale Secret Leak Analysis}Meli et al.~\cite{meli2019bad} scanned billions of GitHub files, achieving high accuracy (99.29\%) with limited regex and entropy filters. This reliance reduced coverage, unlike our work which requires contextual reasoning for the mixed language and code in issue reports.\paragraph{LLM-based Secret Detection}Rahman et al.~\cite{rahman2025secret} proposed a hybrid approach using regex and LLMs for secret classification in source code. This leverages LLMs' contextual understanding to reduce false positives but does not focus on the mixed content of issue reports.\paragraph{Program Analysis for Secret Detection}Sinha et al.~\cite{sinha2015detecting} used program slicing and patterns, achieving high precision ($\sim$100\%) for specific keys (e.g., AWS). While effective for certain codebases, this method is less adaptable to the diverse content in issue reports.\paragraph{Secret Detection Tools}Tools like TruffleHog and Gitleaks use regexes and entropy but often generate false positives and lack contextual understanding for obfuscated or non-standard secrets. Our pipeline enhances regex extraction with language model reasoning to filter out dummy values.\paragraph{Machine Learning-Based Filtering}Saha et al.~\cite{saha2020secrets} used a Soft Voting Classifier with handcrafted features, achieving an F1-score of 86.7\%. Its reliance on hand-engineered features limited generalization. We use transformer-based models to learn semantic and contextual cues directly from raw text, enabling broader application to non-code artifacts.\paragraph{Deep Learning Approaches}Feng et al.~\cite{feng2022automated} used deep neural networks for specific secret types, emphasizing the role of context. Our LLM-based framework covers a broader range (API keys, tokens) and is adapted via contextual fine-tuning for the informal, diverse nature of issue reports.\paragraph{Comparative Analysis of Detection Tools}Basak et al.~\cite{basak2023comparative} compared various open-source and commercial tools, but only on source code. We extend this analysis to issue tracking data, fine-tuning language models for secret detection in unstructured text.\\
Overall, prior work predominantly targeted source code and configuration files. We address the understudied problem of secret leaks in issue reports by leveraging pre-trained language models to expand detection beyond traditional code analysis.
\section{Threats to Validity}
\label{sec:threats}

\noindent \textbf{Internal validity} refers to factors within the study that may affect the results. Our findings could be influenced by model choice and limited hyperparameter tuning due to resource limits. There may also be overlap between our dataset and the pretraining data of some LLMs, slightly biasing performance. To reduce these effects, all models were trained and tested with fixed random seeds, consistent data splits, and reproducibility scripts to ensure stable results.

\noindent \textbf{External validity} concerns how well the results generalize beyond our setup. Since the dataset was built entirely from public GitHub issues, it may not fully reflect proprietary platforms like Jira or GitLab. To improve generalizability, we included diverse repositories, programming languages, and project sizes, and tested performance on 178 unseen repositories to approximate real-world use.

\noindent \textbf{Construct validity} examines whether the study measures what it intends to, which is detecting secrets in issue reports. Labels were created using human judgment and regex-based extraction, which may introduce bias, especially for placeholders or incomplete credentials. To reduce this, we followed detailed annotation rules and measured inter-rater agreement, which showed high reliability. Another potential construct validity threat stems from using regular expressions as a first-layer candidate generator, since secrets not matched by these patterns would never reach the classifier. To evaluate this risk, we manually reviewed 500 randomly sampled issue reports, including ones not flagged by any regex. We found no missed true secrets, suggesting that the regex-based extraction achieves near-complete recall in practice for secrets typically found in GitHub issue reports. This is consistent with prior work showing that regex-based secret scanners favor recall over precision.

\noindent \textbf{Conclusion validity} relates to the soundness of the inferences made from the results. We used standard precision, recall, and F1 metrics, but class imbalance and prompt sensitivity could still affect outcomes. To minimize this, we used stratified data splits, evaluations on held-out test sets, and publicly released all models and code for independent verification and future replication.
\section{Conclusion}
\label{sec:conclusion}

This is the first major study to investigate secret leaks found within software issue reports, addressing a serious but often overlooked aspect of software security. We built a comprehensive benchmark dataset and evaluated both proprietary and open-source language models to separate true secrets from false positives in issue reports. Our hybrid pipeline, which combines regex-based extraction with contextual classification using fine-tuned LLMs, achieved clear gains over traditional methods. Fine-tuned Qwen and LLaMA models achieved 94\% F1-score, while lightweight models such as RoBERTa and CodeBERT also showed strong performance suitable for practical deployment. Future work will focus on enhancing model reasoning ability, adopt retrieval-augmented strategies, improving prompting strategies for better contextual judgment, and expanding the dataset to include more diverse and realistic issue reports.
\bibliographystyle{ACM-Reference-Format}
\bibliography{main}

\end{document}